\documentclass[11pt,a4paper]{iopart}
\usepackage{varioref}
\usepackage[dvips]{graphicx}
\usepackage[latin1]{inputenc}
\usepackage{iopams}

\newcommand{\AmS}{{\protect\the\textfont2
  A\kern-.1667em\lower.5ex\hbox{M}\kern-.125emS}}
\renewcommand{\pt}{\ensuremath{p_{\mathrm{T}}}}
\newcommand{\ks}{\ensuremath{K^0_{\mathrm{s}}}}
\newcommand{\gevc}{\ensuremath{\mathrm{GeV}/c}}

\newcommand{\rcp}{\ensuremath{R_{\mathrm{CP}}}}

\begin{document}

\title[High \pt\ Identified Particles in NA49]{High \pt\ Spectra of Identified Particles Produced in Pb+Pb Collisions at 158$A$ GeV Beam Energy}

\author{Tim Schuster and Andr\'{a}s L\'{a}szl\'{o} for the NA49 Collaboration\footnote[1] {Presented at Strangeness in Quark Matter 2006, Los Angeles, California, USA}}

\author{ \small
C.~Alt$^{9}$, T.~Anticic$^{23}$, B.~Baatar$^{8}$,D.~Barna$^{4}$, 
J.~Bartke$^{6}$, L.~Betev$^{10}$, H.~Bia{\l}\-kowska$^{20}$,
C.~Blume$^{9}$, B.~Boimska$^{20}$, M.~Botje$^{1}$
% ...
}

\vspace{0.5cm}

\address{$^{1}$NIKHEF, Amsterdam, Netherlands.}
% ...
\address{$^{4}$KFKI Research Institute for Particle and Nuclear Physics, Budapest, Hungary.}
% ...
\address{$^{6}$Institute of Nuclear Physics, Cracow, Poland.}
% ...
\address{$^{8}$Joint Institute for Nuclear Research, Dubna, Russia.}
\address{$^{9}$Fachbereich Physik der Universit\"{a}t, Frankfurt, Germany.}
\address{$^{10}$CERN, Geneva, Switzerland.}
% ...
\address{$^{20}$Institute for Nuclear Studies, Warsaw, Poland.}
% ...
\address{$^{23}$Rudjer Boskovic Institute, Zagreb, Croatia.}

\begin{abstract}
Results of the NA49 collaboration on the production of hadrons with large transverse momentum in Pb+Pb collisions at 158$A$ GeV beam energy are presented. A range up to $\pt=4~\gevc$ is covered.
The nuclear modification factor \rcp\ is extracted for $\pi^{\pm}$, $K^{\pm}$ and $p$, and the baryon to meson ratios $p/\pi^+$, $\bar{p}/\pi^-$ and $\Lambda/\ks$ are studied.
All results are compared to other measurements at SPS and RHIC and to theoretical calculations.
\end{abstract}

\submitto{\JPG}


\section{Introduction}
The features of high \pt\ hadron production at RHIC suggest the creation of a new state of matter~\cite{Arse04}--\cite{Adco04}.
While nuclear modification factors smaller than one at high \pt\ are interpreted as a sign for energy loss of high momentum partons in a dense strongly interacting medium, the measured high values in baryon/meson ratios can be explained by hadron production via quark coalescence.
Studying these effects at the SPS can help to enhance the understanding of the underlying mechanisms.
The presented results at the highest SPS beam energy of 158$A$ GeV ($\sqrt{s_{\mathrm{NN}}}=17.2~\mathrm{GeV}$) exhibit similar trends as previously seen in RHIC data.
They thus complement the picture obtained from numerous other observables (like e.g.\ the new results on $\Lambda$ flow~\cite{Kiko06}) indicating that at the top SPS energy a state of matter resembling that at RHIC is created.

\section{Analysis}
The main tracking device of NA49 are four large volume time projection chambers (TPCs). A calorimetric measurement of the projectile spectator energy provides an independent control of the collision centrality. For a detailed description of the detector, see~\cite{Afan99}.

$\pi^{\pm}$, $K^{\pm}$, $p$ and $\bar{p}$ are identified via their specific energy loss in the TPCs ($\mathrm{d}E / \mathrm{d}x$). The resolution of this measurement is 3--6\% and the tracking efficiency is above $95\%$.
In this analysis of charged hadrons, the collision centrality ranges (0--5)\%, (12.5--23.5)\% and (33.5--80)\% of the total inelastic cross section were used. The presented results refer to the rapidity interval $-0.3 < y_{\mathrm{CM}} < 0.7$.
Neutral strange hadrons are identified by a reconstruction of the decay topologies $\ks \rightarrow \pi^{+} \pi^{-}$ ($\mathrm{BR}=68.95\%$) and $\Lambda \rightarrow p \pi^{-}$ (63.9\%). Candidates for these decays are selected by geometrical and kinematical criteria and their invariant mass is calculated to extract yields in $y-\pt$-bins. Here, the centrality interval (0--23.5)\% and the range $-0.5 < y_{\mathrm{CM}} < 0.5$ around midrapidity were used.

All results are corrected for acceptance and reconstruction inefficiency. The $\Lambda$, proton and pion yields have not yet been corrected for feeddown from the decay of heavier particles, but the bias has been estimated to be below 10\%.

\section{Results}

\begin{figure}
\begin{center}
\includegraphics[width=0.9\textwidth]{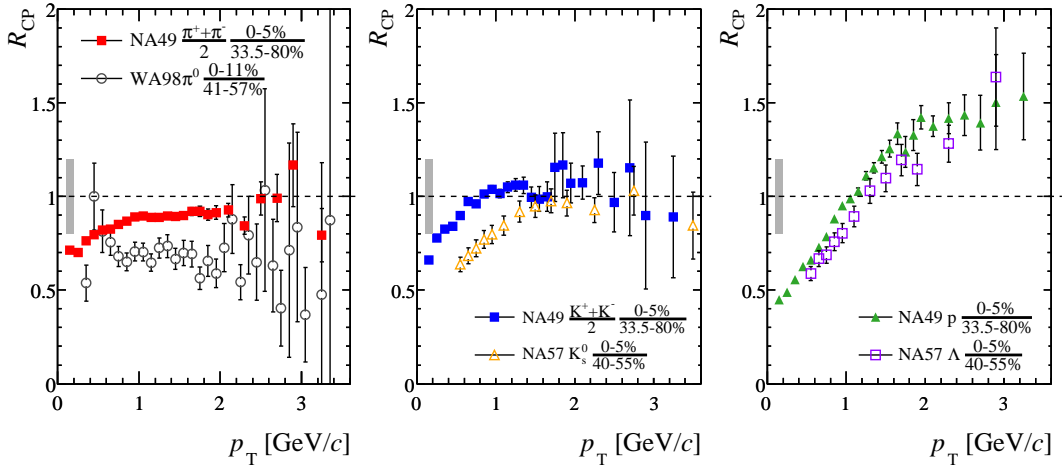}
\end{center}
\caption{A comparison of \rcp\ measurements in Pb+Pb collisions at the top SPS energy of $\sqrt{s_{\mathrm{NN}}}=17.2~\mathrm{GeV}$: $\pi^0$ from WA98 (left panel,~\cite{Agga01}), \ks\ and $\Lambda$ from NA57 (middle and right panels,~\cite{Anti05}), together with the NA49 results. The shaded bands indicate a systematical error on the NA49 results (see text).}
\label{fig:R_CP_SPS}
\end{figure}

The nuclear modification factor \rcp\ is defined as the yield ratio in central over peripheral collisions, scaled by the mean number of binary nucleon-nucleon collisions $\left< N_{\mathrm{coll}} \right>$:
\begin{equation*}
\rcp := \frac{\left< N_{\mathrm{coll}} \right>^{\mathrm{Per.}}}{\left< N_{\mathrm{coll}} \right>^{\mathrm{Cen.}}} \frac{\left( \mathrm{d}^2N / \left( \mathrm{d} \pt \mathrm{d} y \right) \right) ^{\mathrm{Cen.}}}{\left( \mathrm{d}^2N / \left( \mathrm{d} \pt \mathrm{d} y \right) \right) ^{\mathrm{Per.}}}
\end{equation*}
$\left< N_{\mathrm{coll}} \right>$ was calculated for each centrality interval using the VENUS model. The scaling may be model dependent and the corresponding systematic error in \rcp\ is assumed to be of the order of $\pm 20\%$. A shaded band indicates this error in Fig.~\ref{fig:R_CP_SPS}.

\begin{figure}
\begin{center}
\includegraphics[width=0.45\textwidth]{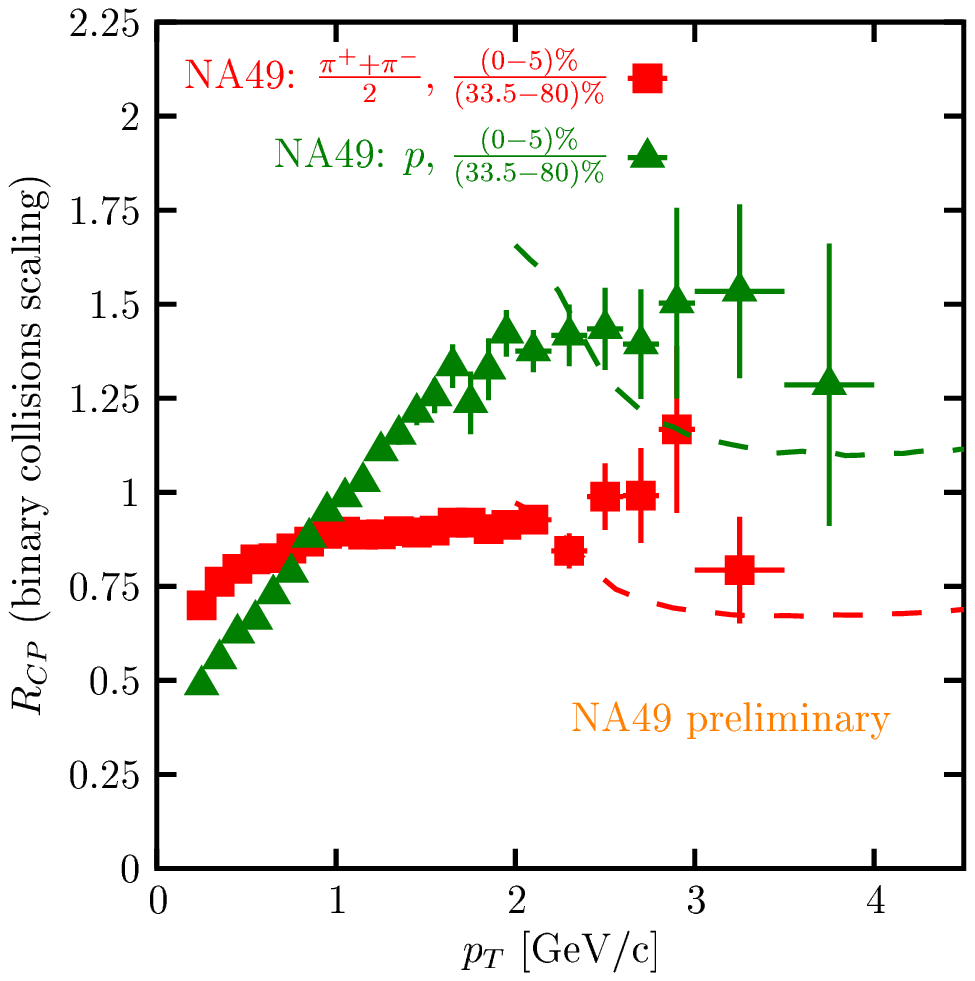}
\includegraphics[width=0.45\textwidth]{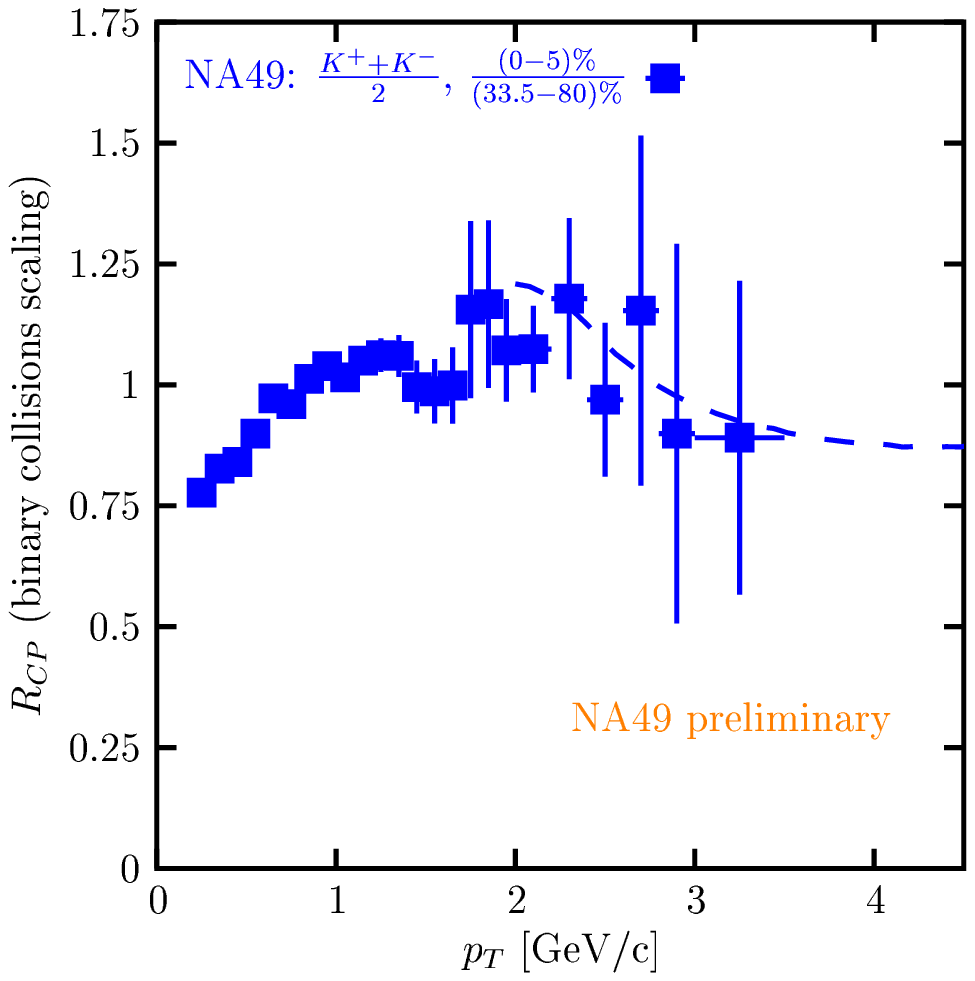}
\end{center} 
\caption{The nuclear modification factor \rcp\ for pions, kaons and protons measured by NA49 in Pb+Pb collisions at $\sqrt{s_{\mathrm{NN}}}=17.2~\mathrm{GeV}$, compared to pQCD model calculations taking into account partonic energy loss (dashed line,~\cite{Wang04}). The compared centrality ranges are indicated as fraction of the total inelastic cross section $\sigma/\sigma_{\mathrm{tot.}}^{\mathrm{inel.}}$.}
\label{fig:R_CP_NA49}
\end{figure} 

\rcp\ for charged pions, kaons and protons is shown in Fig.~\ref{fig:R_CP_NA49}. A monotonous rise leads up to a plateau that is reached at $\pt=1~\gevc$ (for pions and kaons), and at $\pt=2~\gevc$ (for protons) respectively. A perturbative QCD calculation taking radiative energy loss of partons into account~\cite{Wang04} is consistent with the data when considering statistical and systematic errors.
Figure~\ref{fig:R_CP_SPS} shows a comparison to other data available at the highest SPS energy of 158$A$ GeV~\cite{Agga01},\cite{Anti05}.
When considering all systematic errors introduced by different methods of centrality determination and the calculation of $\left< N_{\mathrm{coll}} \right>$, the results of all three experiments are consistent. \rcp\ tends towards the same values at $\pt>2~\gevc$ for the compared hadron species.

\begin{figure}
\begin{center}
\includegraphics[width=0.3\textwidth]{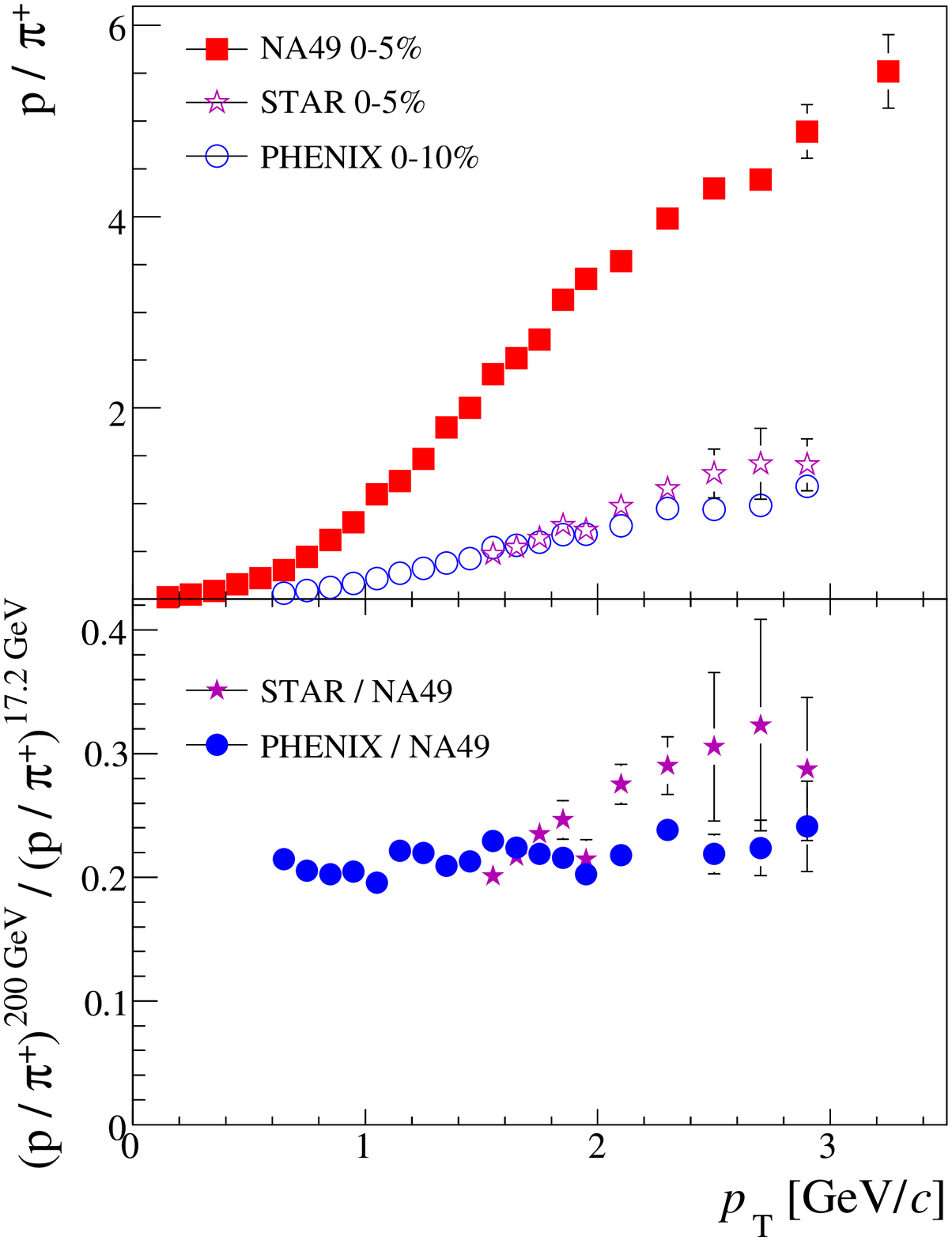}
\includegraphics[width=0.3\textwidth]{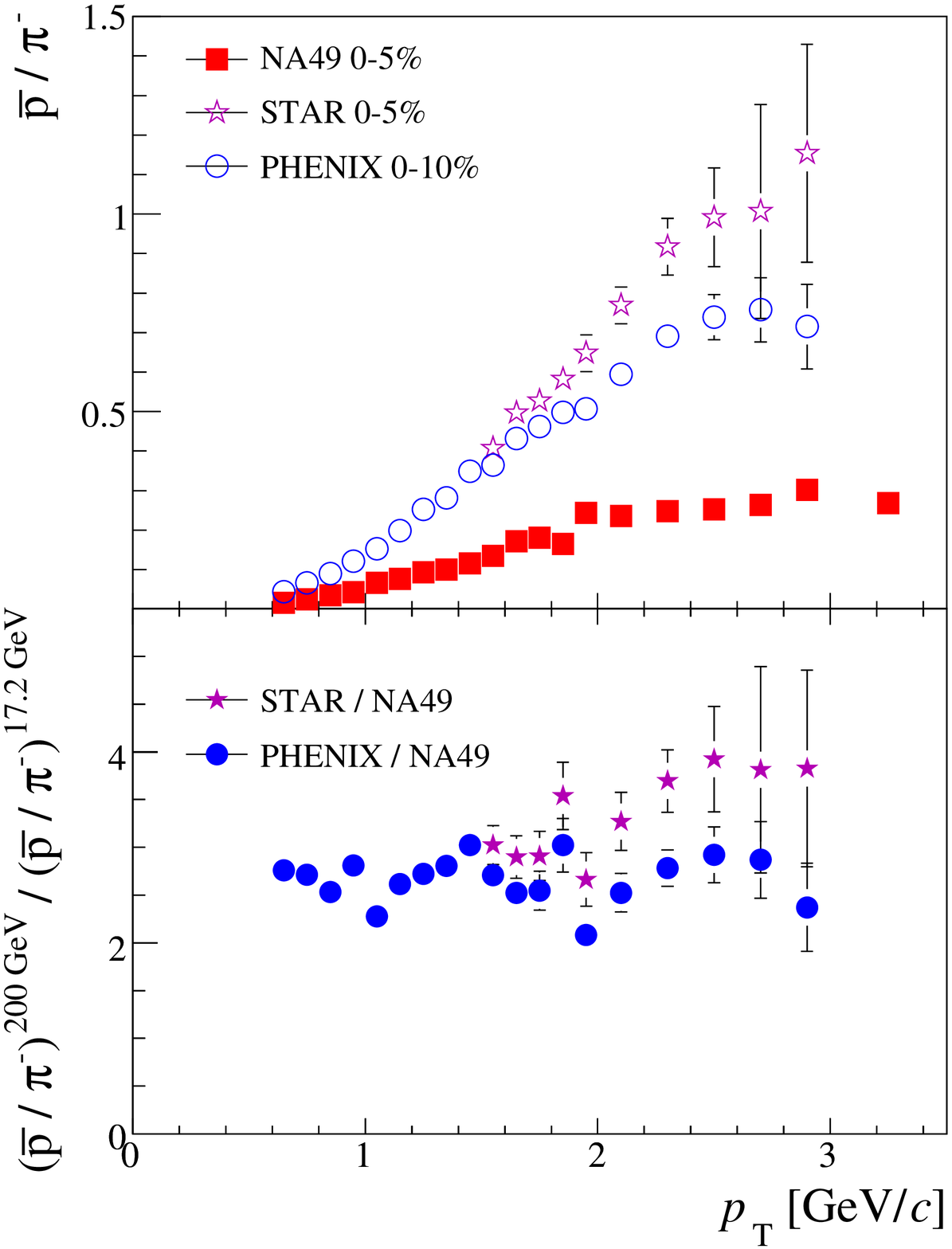}
\includegraphics[width=0.3\textwidth]{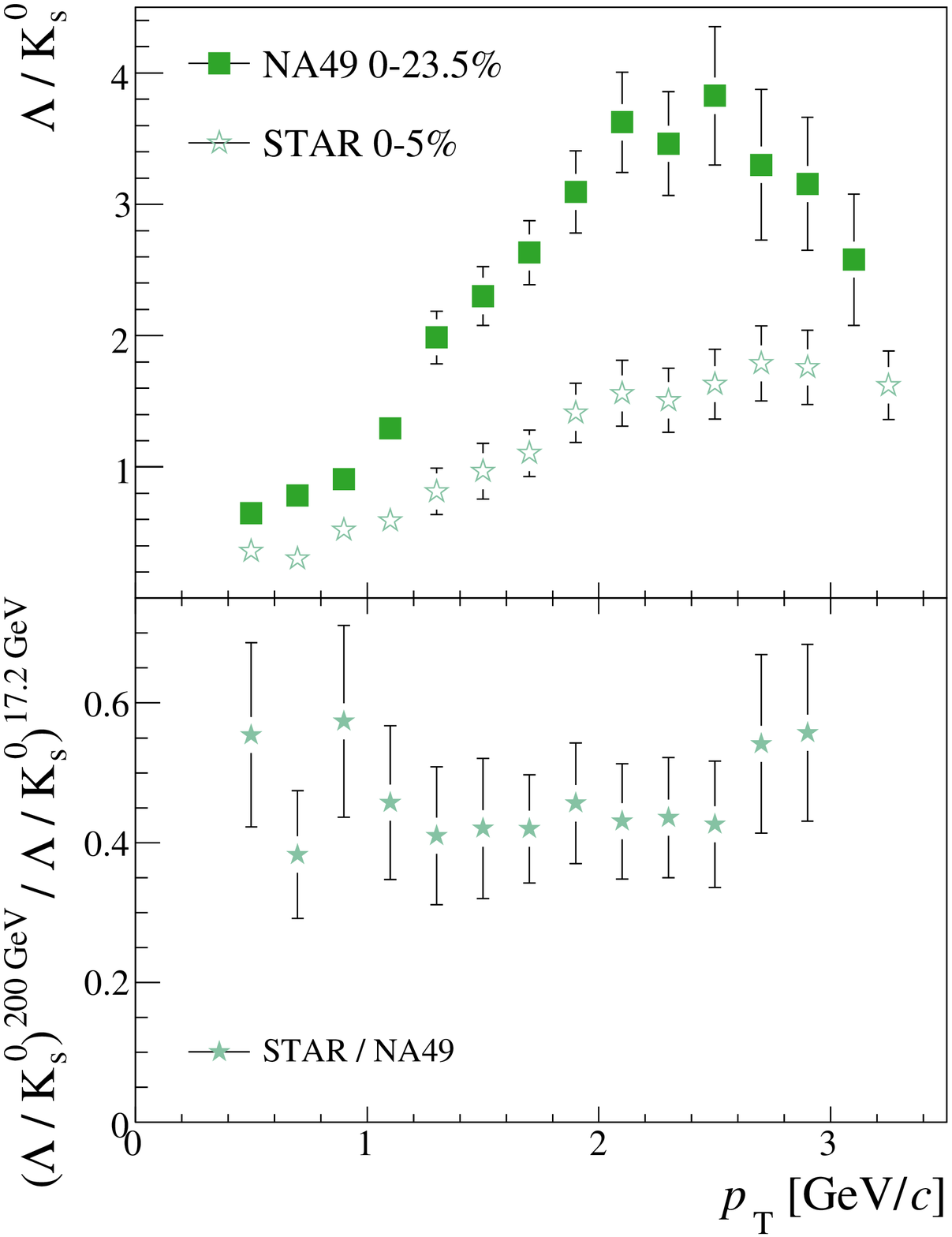}
\end{center}
\caption{The baryon/meson ratios measured in NA49 at $\sqrt{s_{\mathrm{NN}}}=17.2~\mathrm{GeV}$ compared to results from RHIC~\cite{Adam06},\cite{Adle03} at $\sqrt{s_{\mathrm{NN}}}=200~\mathrm{GeV}$. Please note the different scales of the plots.}
\label{fig:Barmes_endep}
\end{figure}

While at RHIC the high values for baryon/meson ratios at intermediate \pt\ can be explained by quark coalescence models (see e.g.~\cite{Adam06}), no calculations of such models are available for SPS energies yet. The upper panels in Fig.~\ref{fig:Barmes_endep} show the NA49 results for $p/\pi^+$, $\bar{p}/\pi^-$ and $\Lambda/\ks$  together with the corresponding results from PHENIX~\cite{Adle03} and STAR~\cite{Adam06} at the highest RHIC energy of $\sqrt{s_{\mathrm{NN}}}=200~\mathrm{GeV}$. The higher net baryon density at SPS energy is manifested in quantitative differences, but the shape of the baryon/meson ratios as a function of \pt\ is the same for both energies. This can be seen in the lower panels of Fig.~\ref{fig:Barmes_endep}, where the double ratios are shown.
New results from RHIC Cu+Cu runs at various energies also show no qualitative change with energy in the shape of the baryon/meson ratios~\cite{Morr06}.

\section{Conclusions}
The presented results on \rcp\ at the highest SPS energy of $\sqrt{s_{\mathrm{NN}}}=17.2~\mathrm{GeV}$ are consistent with a pQCD model that incorporates parton energy loss and that is used to explain RHIC results. Although large systematic errors remain, the results of three different SPS experiments are in line with this picture.
Baryon/meson ratios as a function of \pt\ have the same shape as at RHIC, but the theoretical interpretation in the SPS energy range is still missing.

\section*{Acknowledgments} 
This work was supported by the US Department of Energy
Grant DE-FG03-97ER41020/A000,
the Bundesministerium f\"{u}r Bildung und Forschung, Germany (06F137), 
the Virtual Institute VI-146 of Helmholtz Gemeinschaft, Germany,
the Polish State Committee for Scientific Research (1 P03B 097 29, 1 PO3B 121 29,  2 P03B 04123), 
the Hungarian Scientific Research Foundation (T032648, T032293, T043514),
the Hungarian National Science Foundation, OTKA, (F034707),
the Polish-German Foundation, the Korea Research Foundation Grant (KRF-2003-070-C00015) and the Bulgarian National Science Fund (Ph-09/05).


\section*{References}
\begin{thebibliography}{12}
\bibitem{Arse04} I.~Arsene {\it et al.}  (the BRAHMS Collaboration):
                 Nucl.\ Phys.\ A {\bf 757} (2005) 1. 
\bibitem{Back04} B.~B.~Back {\it et al.}:
                 Nucl.\ Phys.\ A {\bf 757} (2005) 28. 
\bibitem{Adam05} J.~Adams {\it et al.}  (the STAR Collaboration):
                 Nucl.\ Phys.\ A {\bf 757} (2005) 102. 
\bibitem{Adco04} K.~Adcox {\it et al.}  (the PHENIX Collaboration):
                 Nucl.\ Phys.\ A {\bf 757} (2005) 184. 
\bibitem{Kiko06} D.~Kikola {\it et al.} (the NA49 Collaboration):
                 these proceedings.
\bibitem{Afan99} S.~V.~Afanasiev {\it et al} (the NA49 collaboration):
                 Nucl.\ Instrum.\ Meth.\ A {\bf 430} (1999), 210.
\bibitem{Wang04} X.-N.~Wang: Phys.\ Lett.\ B {\bf 595} (2004), 165-170.
\bibitem{Agga01} M.~M.~Aggarwal {\it et al.}  (the WA98 Collaboration):
                 Eur.\ Phys.\ J.\ C {\bf 23} (2002) 225. 
\bibitem{Anti05} F.~Antinori {\it et al.}  (the NA57 Collaboration),
                 Phys.\ Lett.\ B {\bf 623} (2005) 17. 
\bibitem{Adam06} J.~Adams {\it et al.}  (the STAR Collaborations):
                 arXiv:nucl-ex/0601042.
\bibitem{Adle03} S.~S.~Adler {\it et al.}  (the PHENIX Collaboration):
                 Phys.\ Rev.\ C {\bf 69} (2004) 034909. 
\bibitem{Morr06} D.~Morrison {\it et al.} (the PHENIX Collaboration):
                 these proceedings.
\end{thebibliography}
\end{document}